\documentclass[usegraphicx,twocolumn,amsmath,amssymb]{mn2e}

\usepackage{graphicx,amsmath,amssymb}

\title[Relativistic models of magnetized neutron stars]{
Structure and deformations of strongly magnetized neutron stars with twisted torus configurations}

\author[R. Ciolfi {\it et al.}]  {R. Ciolfi\thanks{riccardo.ciolfi@roma1.infn.it}, 
V. Ferrari\thanks{valeria.ferrari@roma1.infn.it}, L. Gualtieri\thanks{leonardo.gualtieri@roma1.infn.it}
  \\
  Dipartimento di Fisica ``G.Marconi'', Sapienza Universit\` a di Roma and Sezione INFN ROMA1, 00185 Roma, Italy }

\begin{document} 
\date{}

\maketitle

\begin{abstract} 
  We construct general relativistic models of stationary, strongly
  magnetized neutron stars. The magnetic field configuration, obtained
  by solving the relativistic Grad-Shafranov equation, is a
  generalization of the {\em twisted torus} model recently proposed in
  the literature; the stellar deformations induced by the magnetic
  field are computed by solving the perturbed Einstein's equations;
  stellar matter is modeled using realistic equations of state. We
  find that in these configurations the poloidal field dominates over
  the toroidal field and that, if the magnetic field is sufficiently
  strong during the first phases of the stellar life, it can produce 
  large deformations.

\end{abstract} 


\section{Introduction}\label{introduction} 
After the discovery of the Soft Gamma Repeaters and Anomalous X-ray
Pulsars \cite{M79,MS95}, a model of these sources was proposed according to
which they are neutron stars with a very strong magnetic field; these {\it magnetars}
would have a surface field as large as $\sim 10^{14}\!-\!10^{15}$ G,
and internal fields about ten times larger \cite{DT}. However, a clear
picture of the structure, dynamics and evolution of magnetars is still
missing. For instance we do not know whether the toroidal components
of the field prevail on poloidal ones and how intense they are.
Consequently, we do not know how large the deformation induced by the
magnetic field on the star is, an information which is essential if
one is interested in the gravitational wave emission of these sources.
A deeper knowledge of the structure of strongly magnetized
neutron stars would also help understanding various astrophysical
processes involving magnetars (intense activity in the X- and gamma-
spectra, quasi-periodic oscillations, eventually gamma-ray bursts).

In a recent paper \cite{C09}, to be referred to as Paper I hereafter,
we constructed stationary models of non rotating neutron stars endowed
with a strong magnetic field, in the framework of General Relativity
(GR). In these models the poloidal field extends
throughout the star and in the exterior, whereas the toroidal field is
confined into a torus-shaped region inside the star, where the field
lines are closed. It is worth reminding that these {\it twisted torus}
configurations have been found to be a quite general outcome of
dynamical simulations of the early evolution of magnetized stars, in
the framework of Newtonian gravity. Furthermore, due to magnetic
helicity conservation, they appear to be stable on dynamical
time-scales \cite{BS0,BN,BS}, are not significantly affected by
rotation \cite{Y06}, and do not depend on the initial angle between
the rotation and the magnetic axes \cite{GR06}.

%
In Lander \& Jones \shortcite{LJ} twisted torus configurations were studied 
in Newtonian gravity; the maximal relative strenght of the toroidal and poloidal
components and the induced stellar deformation were evaluated
using a polytropic equation of state (EOS) to model neutron star matter.

In Paper I, we studied the twisted torus configurations in GR
using a more realistic EOS.
We considered  a
relation between the poloidal and the toroidal components of the field
which is linear in the flux function, and estimated their ratio by
determining the configuration of minimal energy at fixed magnetic
helicity, under the assumption that the contribution of the $l>1$
multipoles is minimum outside the star.

In the present paper, we reconsider the above assumptions: the higher
multipoles contribution is not assumed {\it a priori} to be minimum
outside the star, and we allow for a more general parametrization of
the relation between toroidal and poloidal fields. We determine the
configuration of minimal energy at fixed magnetic helicity, and
evaluate the stellar deformations induced by the twisted torus field by
solving the perturbed Einstein equations including all relevant
higher order multipoles.

As in Paper I,
the magnetized fluid is described in the framework of ideal MHD,
which is accurate only in the first few hours of the star life, when
the crust is still liquid and the matter in the core has not yet
undergone a phase transition to the superfluid state. 
Since the characteristic Alfv\`en time is of the order of $\sim 0.01\!-\!10$
s, the magnetized fluid could reach a stationary state while the
matter is still liquid and not yet superfluid\footnote{We remark that
even in presence of a stable stratification of the chemical
composition, a magnetic field as strong as $B\gtrsim
10^{15}$ is still allowed to evolve throughout the star on a
dynamical time-scale \cite{TM01}.}. The magnetic field induces
quadrupolar deformation on the stellar shape and, as we shall later show, 
for magnetic fields as high as those observed in magnetars 
this deformation would be large; when  the crust forms, it
would maintain this deformed shape.
The magnetic field would
subsequently evolve on time-scales of the order of $\sim10^3\!-\!10^5$
years due to dissipative effects like ohmic decay, ambipolar diffusion and
Hall drift \cite{GR92,WT,PG07}.

The structure of the paper is as follows. In Section \ref{TTMFC} we
generalize the twisted torus magnetic field configurations introduced
in Paper I by dropping the assumption that the contribution from the
$l>1$ multipoles outside the star is minimum, and using a more general
parametrization of the functional relation between toroidal and
poloidal fields. In Section \ref{ell} we determine the stellar
deformations induced by the magnetic field. In Section
\ref{conclusions} we draw our conclusions.

\section{Twisted torus magnetic field configuration}\label{TTMFC}
In this Section
we briefly describe the formalism and the basic equations we solve to
determine the twisted torus magnetic field configuration; furthermore, we
discuss  the
modifications introduced with respect to the analysis carried out in
Paper I.

\subsection{The model}\label{model}
We assume that the magnetized star is non-rotating, stationary and
axisymmetric. 
The magnetized fluid is described in the framework of ideal MHD, in which
the effects of electrical conductivity are neglected.
Furthermore, we assume a vacuum exterior.
We follow the same notation and conventions as in Paper I.
We treat the magnetic field as a stationary, axisymmetric perturbation
of a spherically symmetric background with metric
\begin{equation}
ds^2=-e^{\nu(r)}dt^2+e^{\lambda(r)}dr^2+r^2(d\theta^2+\sin^2\theta d\phi^2)~,
\end{equation}
where $\nu(r)$ and $\lambda(r)$ are solutions of the unperturbed Einstein
equations; the unperturbed four-velocity is
$u^\mu=(e^{-\nu/2},0,0,0)$. 

To model neutron star matter in the core
we use the Akmal, Pandharipande, Ravenhall EOS, named APR2 \cite{APR},
and the Glendenning EOS named GNH3 \cite{G85};
the crust is modeled using a standard EOS
which accounts for the density-pressure relation in the
crustal region, but not for its elastic properties 
(see Benhar, Ferrari \& Gualtieri \shortcite{BFG}).
For a neutron star with mass $M=1.4~M_\odot$, the APR2 star
has a large compactness ($R=11.58$ km),
whereas the compactness of the GNH3 star is
small ($R=14.19$ km).

As shown in Colaiuda {\it et al.} \shortcite{C08}, an appropriate
gauge choice allows to write the potential $A_\mu$, in terms of which
the Maxwell tensor is written ($F_{\mu\nu}=\partial_\nu
A_\mu- \partial_\mu A_\nu$), as
\begin{equation}
A_\mu=\left(0,e^{\frac{\lambda-\nu}{2}}\Sigma,0,\psi\right)\,;
\end{equation}
the two functions $\Sigma(r,\theta)$  and $\psi(r,\theta)$
describe the toroidal and the poloidal field,
respectively. Here $A_\mu$ is considered as a  first order quantity, $O(B)$.
Furthermore, the $\phi$-component of Euler's equation
yields $-\psi_{,r}J^r-\psi_{,\theta}J^\theta=O(B^4)$. Using Maxwell's
equations and neglecting higher-order terms one finds the
integrability condition
\begin{equation}
\left(\sin\theta\Sigma_{,\theta}\right)_{,\theta}\psi_{,r}-\left(\sin\theta\Sigma_{,\theta}\right)_{,r}\psi_{,\theta}=0~,
\end{equation}
which implies that $\sin\theta\Sigma_{,\theta}$ is a function of $\psi$:
\begin{equation}
\sin\theta\Sigma_{,\theta}\equiv\beta(\psi)=\zeta(\psi)\psi\,.\label{defbeta}
\end{equation}
The function $\zeta(\psi)=\beta(\psi)/\psi$ represents the ratio
between the toroidal and poloidal components of the magnetic field and
characterizes the kind of field configuration we want to model. For
instance, configurations with $\zeta=$ constant have been studied in
Ioka \& Sasaki \shortcite{IS}, Colaiuda {\it et al.} \shortcite{C08},
Haskell {\it et al.} \shortcite{Haskell}. If
the space outside the  star is assumed to be vacuum, the toroidal field, and consequently
$\zeta$, must vanish for $r > R$, where $R$ is the neutron star radius; 
in this case, the choice $\zeta=$ constant yields an inconsistency, 
unless one assumes that surface
currents cancel the toroidal field outside the star, or imposes that
the constant $\zeta$ assumes very particular values.

These problems do not arise with the twisted torus configurations (see
for instance Paper I or Lander \& Jones \shortcite{LJ}, Yoshida {\it
  et al.} \shortcite{Y06}) since the toroidal field is confined in a
region inside the neutron star, 
and the magnetic field is continuous everywhere.
For these configurations
the function $\beta(\psi)$ is continuous, and  has the form
\begin{equation}
\beta(\psi)\sim\Theta(|\psi/\bar\psi|-1)\,,\label{gentwth}
\end{equation}
where $\bar\psi\equiv\psi(R,\pi/2)$ is the value of  function $\psi$,
which describes the  poloidal field, on the stellar surface,
and $\Theta$ is the Heaviside step function. 
If  $\beta(\psi)$ satisfies Eq.~(\ref{gentwth}), the magnetic field
\begin{equation}
B^\mu=\left[0,\frac{e^{-\frac{\lambda}{2}}}{r^2\sin{\theta}}\psi_{,\theta} \,,
-\frac{e^{-\frac{\lambda}{2}}}{r^2\sin{\theta}}\psi_{,r} \,, 
-\frac{e^{-\frac{\nu}{2}} \beta(\psi)}{r^2\sin^2{\theta}}\right]
\label{genB}
\end{equation}
for $r>R$ becomes purely poloidal, consistently
with the assumption of vacuum outside the star. 

To find the field configuration we need to solve 
the relativistic Grad-Shafranov (GS) equation, which
follows from Euler's and Maxwell's equations (see Paper I for details)
and has the form
\begin{eqnarray}
&&-\frac{e^{-\lambda}}{4\pi}\left[\psi''+\frac{\nu'-\lambda'}{2}\psi'
\right] -\frac{1}{4\pi r^2}\left[\psi_{,\theta\theta}-\cot{\theta}
\psi_{,\theta} \right] 
\nonumber\\
&&-\frac{e^{-\nu}}{4\pi}\beta\frac{d\beta}{d\psi}=(\rho+P)r^2\sin^2{\theta} [c_0+c_1\psi] \;\; .
\label{genGS}
\end{eqnarray}
The constants $c_0$, $c_1$ characterize the $\phi$-component of the
current density inside the star, which has the form
\begin{equation}
J_\phi=\frac{e^{-\nu}}{4\pi}\beta\frac{d\beta}{d\psi}+(\rho+P)r^2\sin^2\theta[c_0+c_1\psi+O(B^2)]\,.\label{defc0c1}
\end{equation}
\begin{figure*}
\centering
\begin{center}
\includegraphics[width=5cm,angle=270]{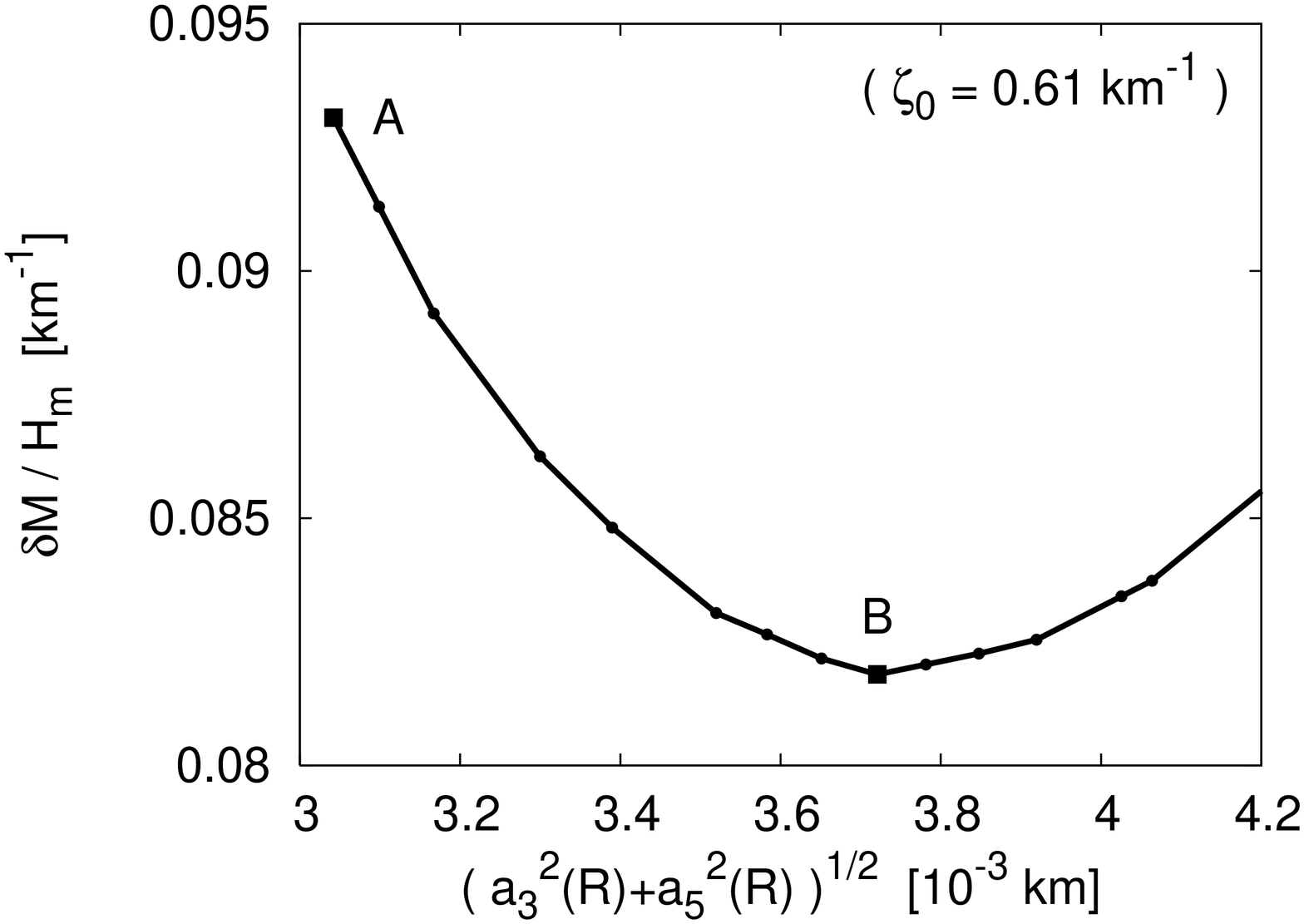}
\includegraphics[width=5cm,angle=270]{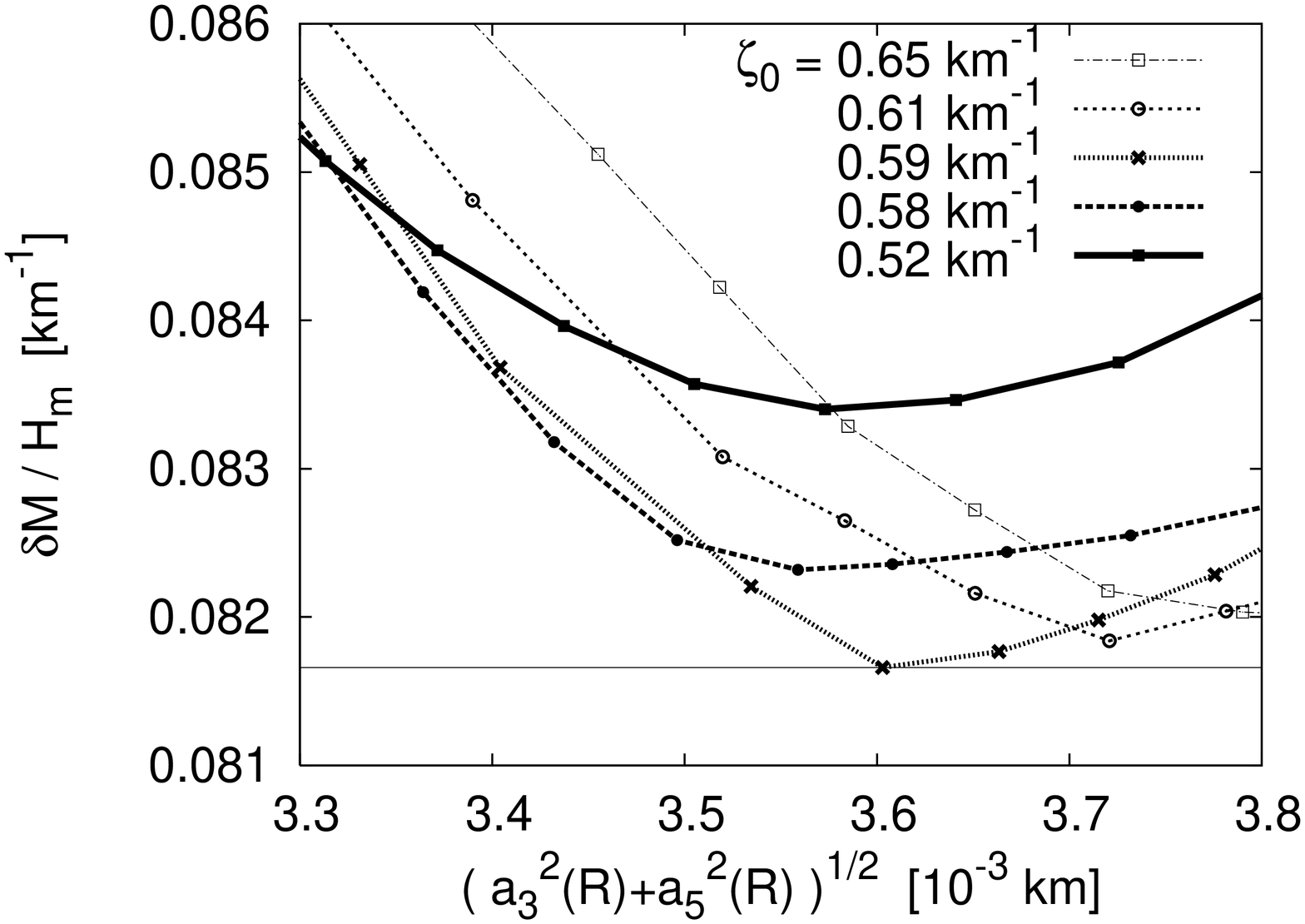}
\end{center} 
\caption{The function $\delta M/H_m$ is plotted as a 
function of $\sqrt{a_3^2(R)+a_5^2(R)}$;   
on the left $\zeta_0=0.61$ km$^{-1}$, on the right the cases 
$\zeta_0=0.65$, $0.61$, $0.59$, $0.58$ and $0.52$ km$^{-1}$   
are shown together for comparison.
\label{HM}}
\end{figure*}
If we now define
$\psi(r,\theta)\equiv \sin{\theta}a(r,\theta)_{,\theta}$, expand
the function $a(r,\theta)$ in Legendre polynomials
\begin{equation}
a(r,\theta)=\sum_{l=1}^{\infty} a_l(r)P_l(\cos{\theta})
\label{expa}
\end{equation}
and project the GS equation onto the different harmonic components, we
find a system of coupled ordinary differential equations for the
functions $a_l(r)$. 
These equations  are solved by imposing
the following boundary conditions. (i) The functions
$a_l(r)$ have a regular behaviour at the origin; an asymptotic
expansion of the GS equation shows that this implies
\begin{equation}
a_l(r\rightarrow 0)=\alpha_l r^{l+1}\,.
\label{alphal}
\end{equation}
(ii) The functions $a_l(r)$ and their derivatives $a'_l(r)$ are
continuous across the stellar surface where they match with the
solutions in vacuum, which are known in an analytical form (see Paper I);
therefore, the ratios $a'_l/a_l$ computed at $r=R$ in
terms of the interior and exterior numerical solutions must
coincide.   (iii) The overall normalization of the field is fixed by
requiring the $l=1$ component of the magnetic field at the pole to be
$B_{pole}=10^{16}$ G; this corresponds to $a_1(R)=1.93\times10^{-2}$
km.

Once the form of the function $\beta(\psi)$  and 
the number $n$ of multipoles we want to include have
been assigned, the field is determined when
we fix $n+2$ arbitrary constants:
the $n$ constants $\alpha_l$ and the two constants
$c_0$, $c_1$ defined in Eq. (\ref{defc0c1}).
$n+1$ of them  are determined by imposing the boundary conditions.
In Paper I the last constant was fixed assuming that the contribution
of  higher-order multipoles outside the star is minimum, i.e. by
minimizing the function $(\sum_{l>1} a_l^2 )/a_1^2$, for $r \ge R$.
In this paper we remove this assumption, thus exploring a
larger parameter space, and fix the last constant by finding
the configurations which are energetically favoured (this is discussed
in detail in the next Section).

To this purpose, we minimize the total energy at fixed magnetic
helicity.  As shown in Paper I, this is equivalent to minimize the
ratio $\delta M/H_m$, where $H_m$ is the magnetic helicity
\begin{equation}
H_m=\frac{1}{2}\int d^3x\sqrt{-g}\epsilon^{0\beta\gamma\delta}F_{\gamma
\delta}A_\beta\,,
\label{helic}
\end{equation}
and $\delta M$ is the mass-energy increase due to the perturbation that
the magnetic field induces on the spherical star. This quantity
is determined in terms of the functions which characterize the
far field limit of the spacetime metric.
Finally, we compute the ratio of the magnetic energy stored in the
poloidal field to the total magnetic energy, $E_p/E_m$.  The equations
to determine these quantities are given in appendix \ref{HME}.

The GS system of equations admits
two particular classes of solutions: the symmetric (with respect to
the equatorial plane) solutions, with vanishing even-order components
($a_{2l}\equiv0$), and the antisymmetric solutions, with vanishing
odd-order components ($a_{2l+1}\equiv0$).
In Paper I the choice of minimizing the $l>1$ multipole contribution
led naturally to symmetric solutions, i.e. those with $a_{2l}\equiv 0$.
It is worth noting that even-order multipoles contribute to the
energy but not to the magnetic helicity; therefore, any solution
minimizing energy at fixed magnetic helicity corresponds to a vanishing
antisymmetric component.
Since in this paper we still look for minimal energy configurations,
we shall consider only symmetric solutions. 
In addition, as in Paper I, we shall restrict to multipoles with $l\le 5$
(the relevance of $l>5$ multipoles is discussed in Paper I, Section 5.3).
\subsection{Relative strength of different multipoles}\label{RSODM} 
\begin{figure*}
\centering
\begin{minipage}{176mm}
\begin{center}
\includegraphics[width=4.7cm,angle=270]{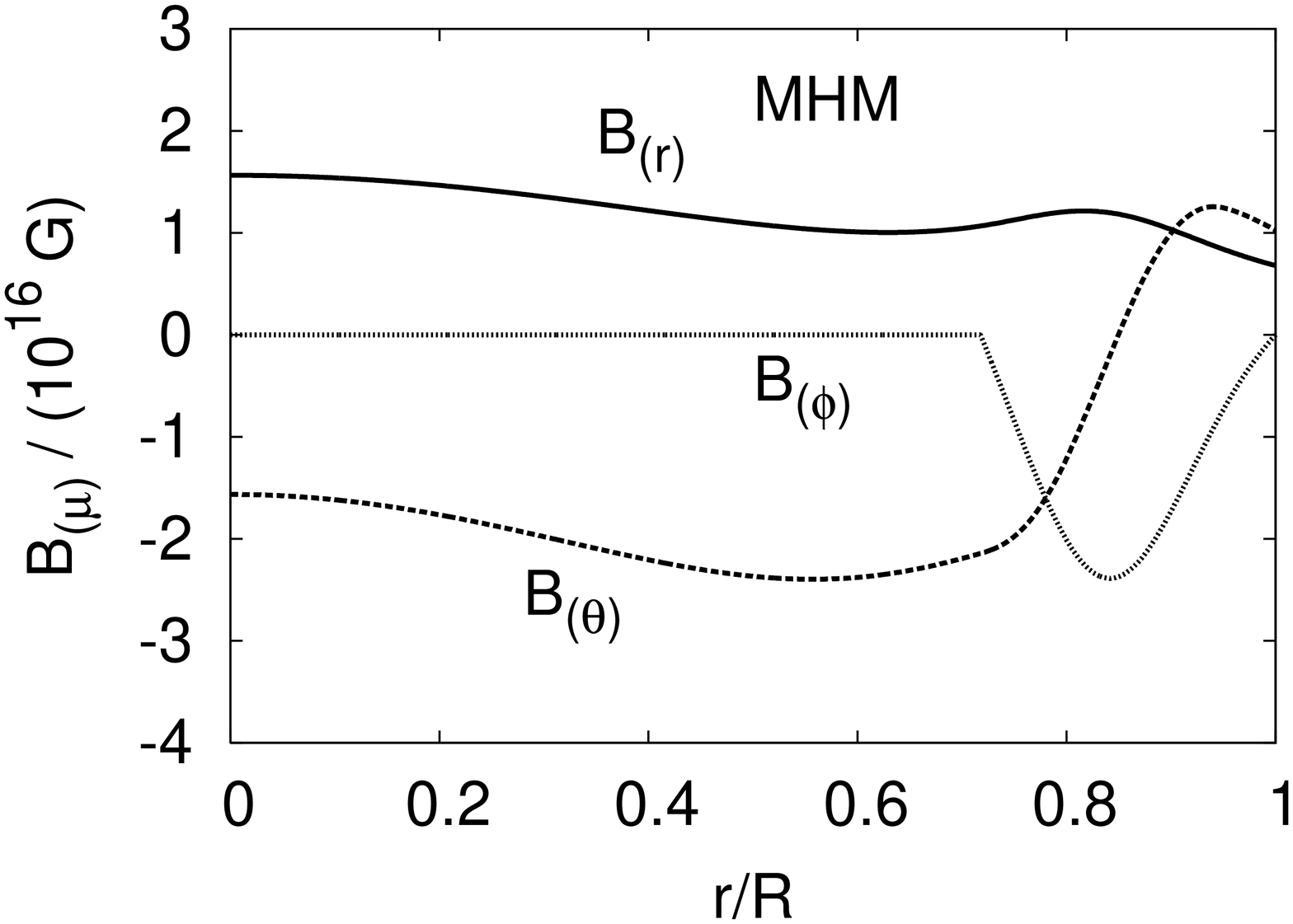}      
\includegraphics[width=4.7cm,angle=270]{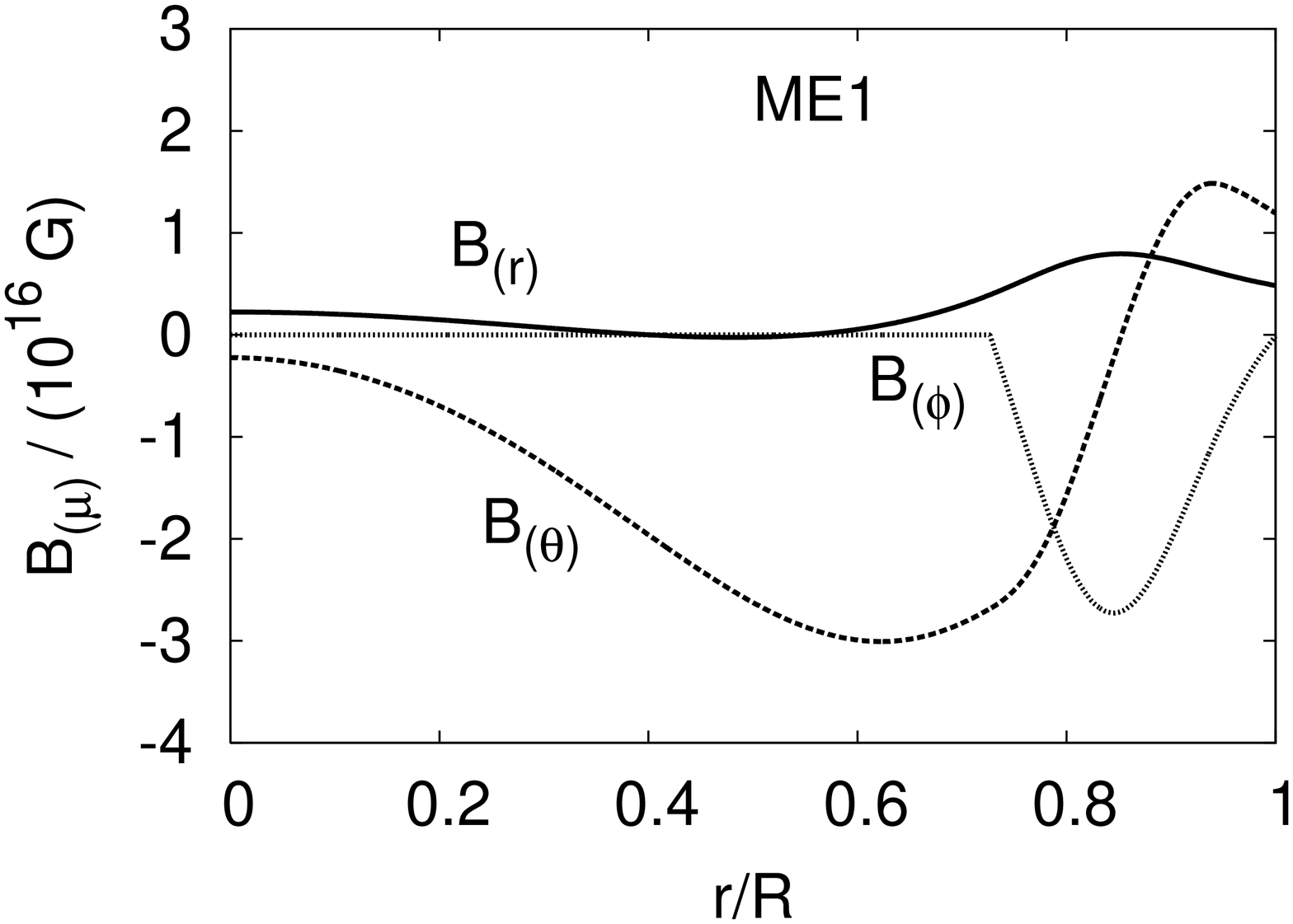}
\end{center} 
\caption{The profiles of the tetrad components of the magnetic field
  $B_{(r)}(\theta=0)$, $B_{(\theta)}(\theta={\pi}/{2})$, $B_{(\phi)}(\theta={\pi}/{2})$
are plotted as functions of the radial distance normalized to the stellar
radius.  The left panel refers to the MHM configuration (energy is minimized
assuming that the contribution of the multipoles higher than $l=1$ 
is minimum for $r >R$); in this case $\zeta_0=0.61$ km$^{-1}$. 
The right panel refers to the minimal energy configuration ME1, obtained
with no assumption on the relative strengths of the
different multipoles, and for $\zeta_0=0.59$ km$^{-1}$.
\label{comparHM}}
\end{minipage} 
\end{figure*}
As explained in the previous section,
since we remove the condition of minimal contribution from higher order
multipoles, the boundary conditions  are not sufficient
to fix all the parameters of the problem and we are left with a free arbitrary
constant. We choose $c_1$ as a ``free'' parameter.

In this section we shall choose the function $\beta(\psi)$
as in Paper I: 
\begin{equation}
\beta(\psi)=\psi \zeta(\psi)=\psi \zeta_0 \left( |\psi/\bar{\psi}|-1 \right) 
\cdot\Theta(|\psi/\bar{\psi}|-1)   \,,
\label{zetaOLD} 
\end{equation}
where $\zeta_0$ is a real parameter. In the next section  we shall consider a
more general form of $\beta(\psi)$ compatible with Eq.~(\ref{gentwth}).
The constant $\zeta_0$
determines the ratio between the amplitudes of the toroidal and poloidal
fields.
Thus, we minimize the energy with respect to two parameters: $c_1$ and
$\zeta_0$. We proceed as follows. 
For assigned values of $\zeta_0$:
\begin{itemize}
\item
we solve the equations for the $a_l$'s  for
different values of $c_1$
\item
we compute $\delta M/H_m$ -- the quantity to minimize --
for the corresponding configurations
\item
we compute the quantity
$\sqrt{a_3^2(R)+a_5^2(R)}$, which represents the surface
contribution of the multipoles higher than $l=1$.
\end{itemize}

In Fig.~\ref{HM} we plot the ratio $\delta M/H_m$ as a function of
$\sqrt{a_3^2(R)+a_5^2(R)}$. In the left panel we fix $\zeta_0=0.61$
km$^{-1}$. In Paper I we showed that, under the assumption of
minimal contribution of higher order multipoles, i.e. 
\[
\sqrt{a_3^2+a_5^2}\qquad\hbox{ minimum for }\quad r \ge R~,
\]
(to hereafter, this will be named the
Minimum High Multipole (MHM) condition),
the quantity $\delta M/H_m$ is minimum 
for this value of $\zeta_0$. This MHM configuration
corresponds to the point $A$ on the curve plotted in Fig.~\ref{HM}.

Since we now drop the MHM condition, the minimum of
$\delta M/H_m$ occurs for a different value of
$\sqrt{a_3^2(R)+a_5^2(R)}$ (point $B$ in Fig.~\ref{HM}), which
corresponds to the energetically favoured configuration for
$\zeta_0=0.61$ km$^{-1}$. For an assigned value of $H_m$, the relative
variation of the total energy of the configuration $B$ with respect to
$A$ is of the order of 13\%. Fig.~\ref{HM} refers to a star with EOS
APR2. Similar results are obtained for the GNH3 star.

In the right panel of Fig.~\ref{HM} we plot $\delta M/H_m$ for 
selected values of $\zeta_0$,
and compare the different profiles. We have explored the parameter space
$(\zeta_0,\sqrt{a_3^2(R)+a_5^2(R)})$ (equivalent to $(\zeta_0,c_1)$), 
finding that the function ${\delta M}/{H_m}$ has a minimum ($\delta
M/H_m=0.0817$) for $\zeta_0=0.59$ km$^{-1}$ and 
$\sqrt{a_3^2(R)+a_5^2(R)}=3.6\times 10^{-3}$ km. It is worth  reminding that
the $l=1$ contribution is 
$a_1(R)=1.93\times10^{-2}$ km. 
We shall refer to this configuration as the Minimal Energy 
1 (ME1) configuration.

In Fig.~\ref{comparHM} we compare the profiles of the tetrad 
components of the magnetic field for the MHM and the ME1  configurations.
We see that, whereas for the MHM configuration $B_{(\theta)}$ and
$B_{(r)}$ are significantly different from zero throughout the star,
for the ME1 configuration,  obtained
with no assumption on the relative strengths of the
different multipoles for $r >R$,
these field components are strongly reduced near the axis.
Conversely, the toroidal component  $B_{(\phi)}$ has a similar behaviour in
both configurations.
The two panels of Fig.~\ref{comparHM} illustrate how the  magnetic
field rearranges inside the star when  the MHM condition is removed. 
The situation can be explained as follows.
The  magnetic helicity $H_m$ can be written as
\begin{equation}
  H_m=-2\pi\int_0^R dr \int_0^\pi 
  (A_r \psi_{,\theta}-\psi A_{r,\theta}) d\theta
 \;\; ; 
\label{firstHm}
\end{equation}
therefore, $H_m$ vanishes if either $\psi=0$, i.e. the poloidal field
vanishes, or $A_r=0$, i.e. the toroidal field vanishes. In the twisted
torus model the toroidal field is zero in the inner part of the star,
therefore $H_m$ receives contributions only from the magnetic field in
the region where $B_{(\phi)}\neq 0$. Since in that region the 
field components of the MHM and ME1 configurations are similar, these
configurations have nearly the same magnetic helicity $H_m$. On the
other hand, the energy $\delta M$ receives contributions from the
field components throughout the entire star, and these
contributions are not vanishing in the region where $B_{(\phi)}= 0$.
\begin{figure*}
\centering
\begin{minipage}{176mm}
\begin{center}
\includegraphics[width=4.7cm,angle=270]{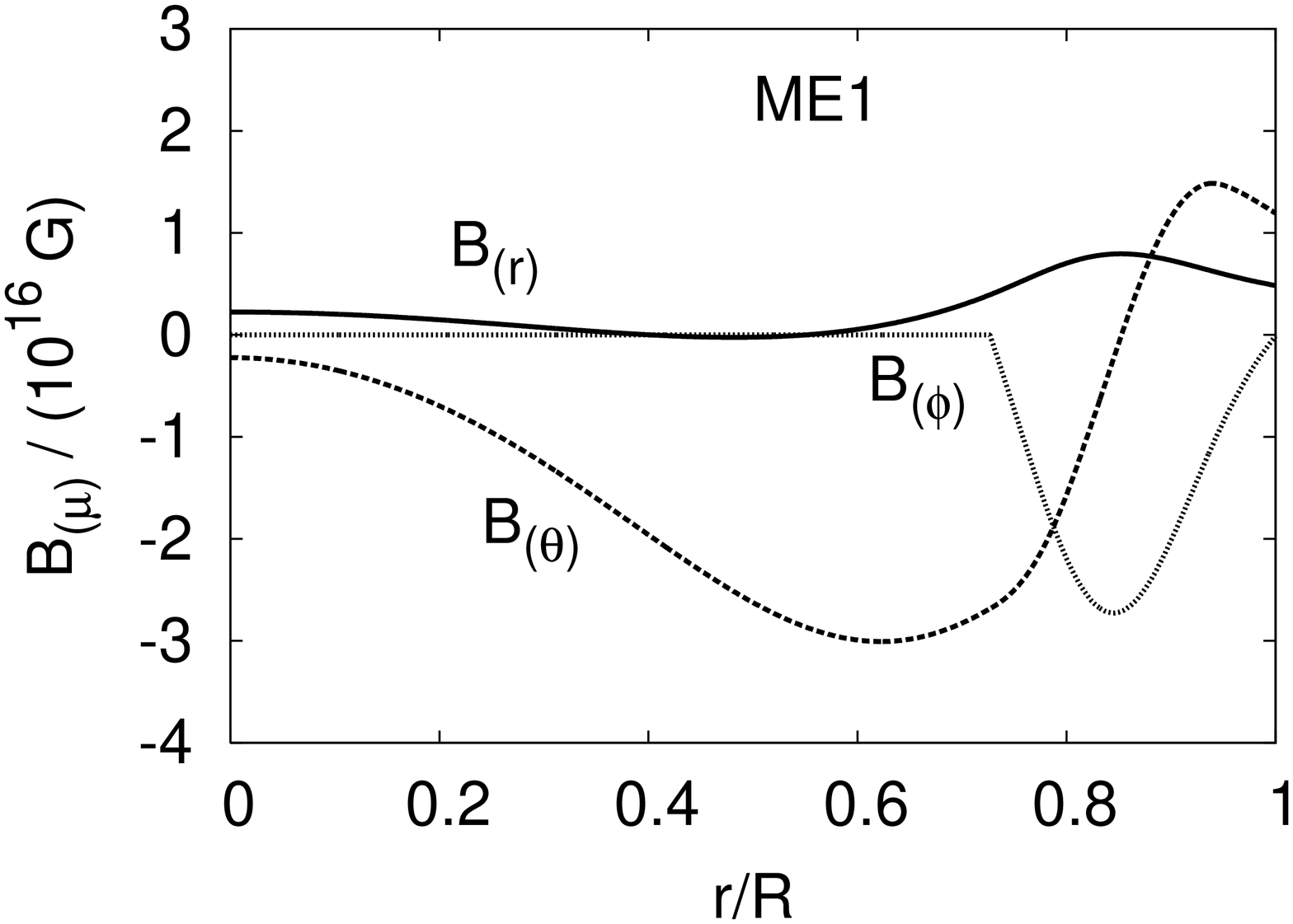}
\includegraphics[width=4.7cm,angle=270]{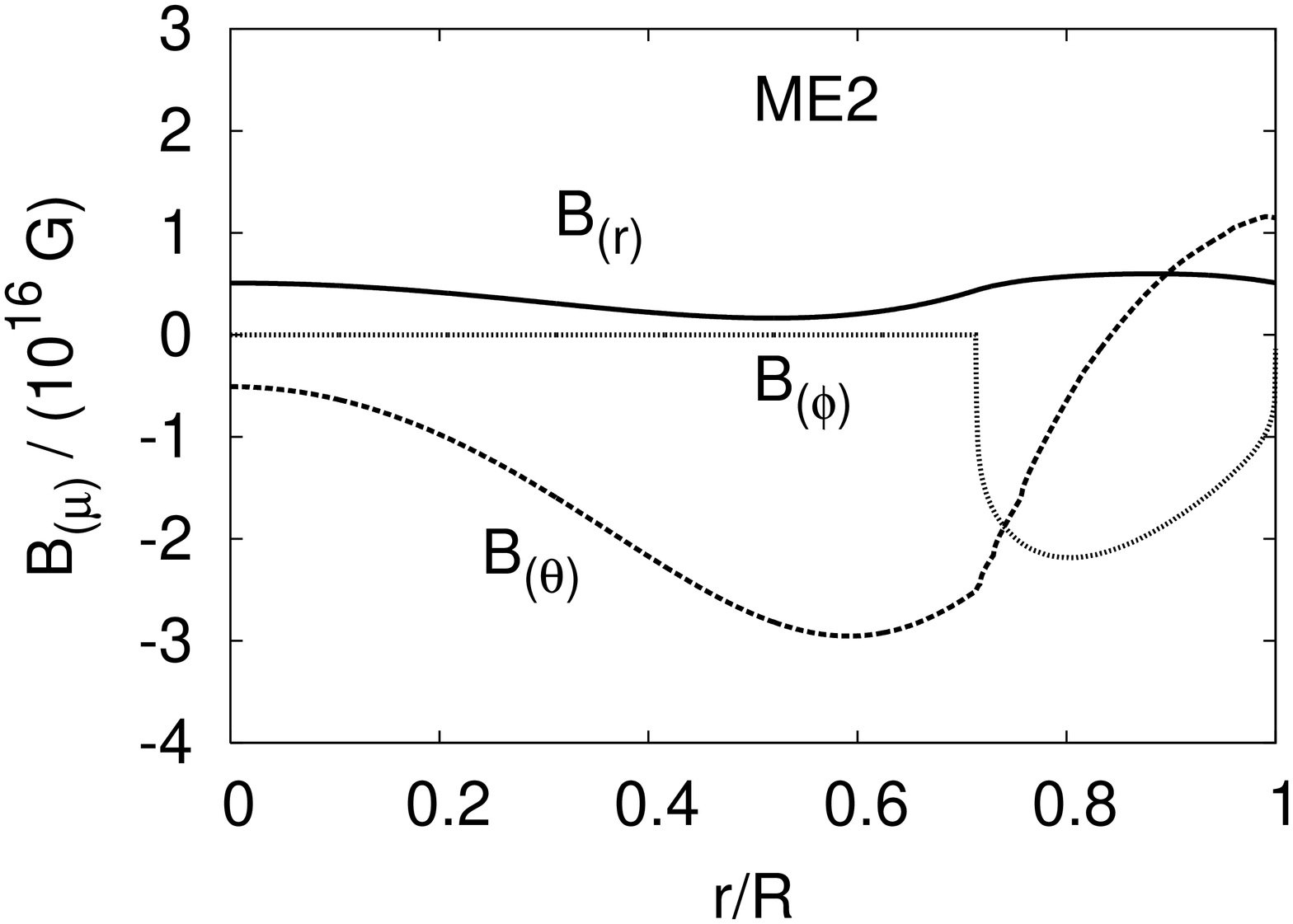}
\end{center}
\begin{center}
\qquad\qquad\qquad\qquad\qquad\includegraphics[width=5.9cm,angle=0]{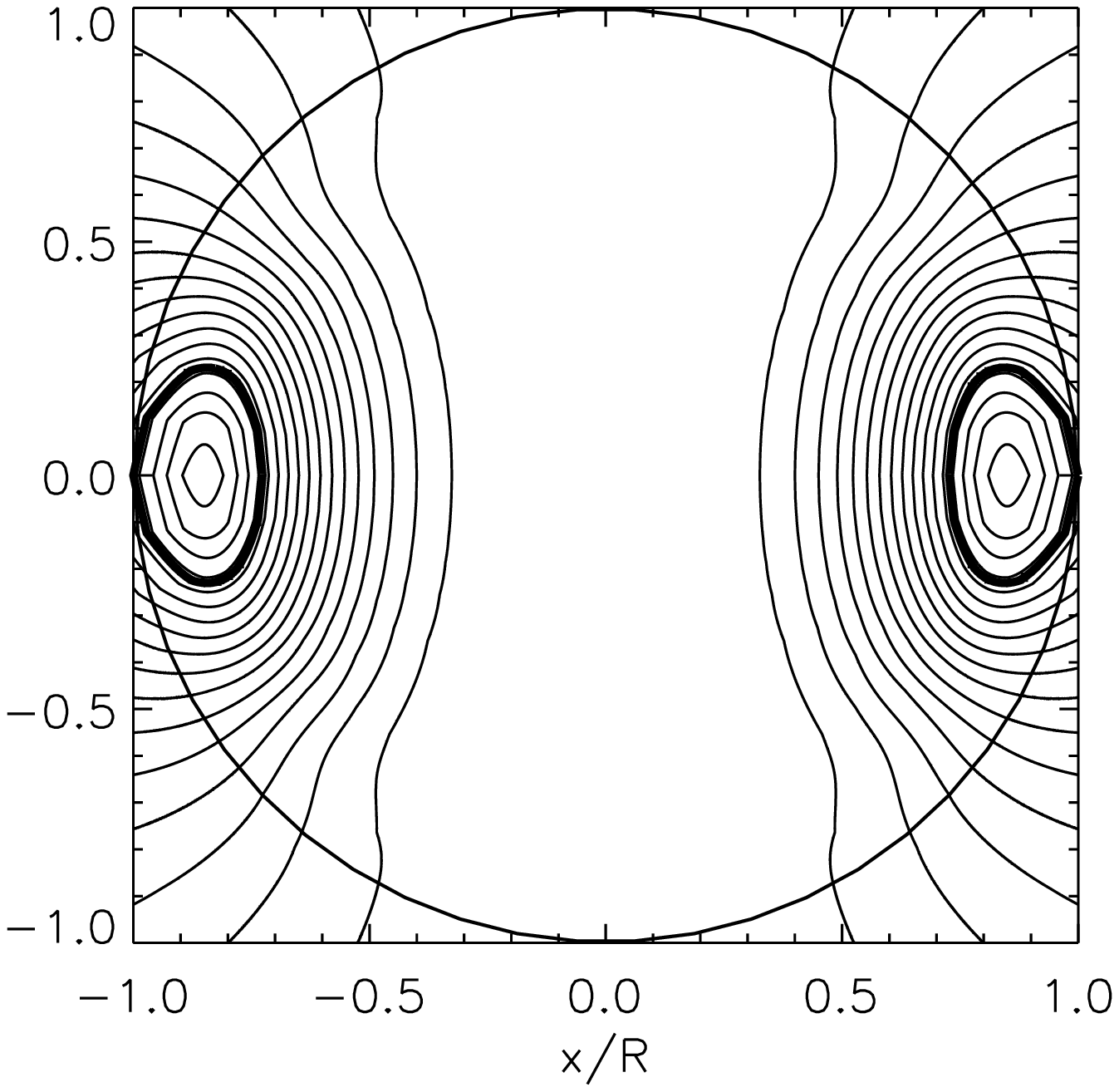}
\qquad\includegraphics[width=5.9cm,angle=0]{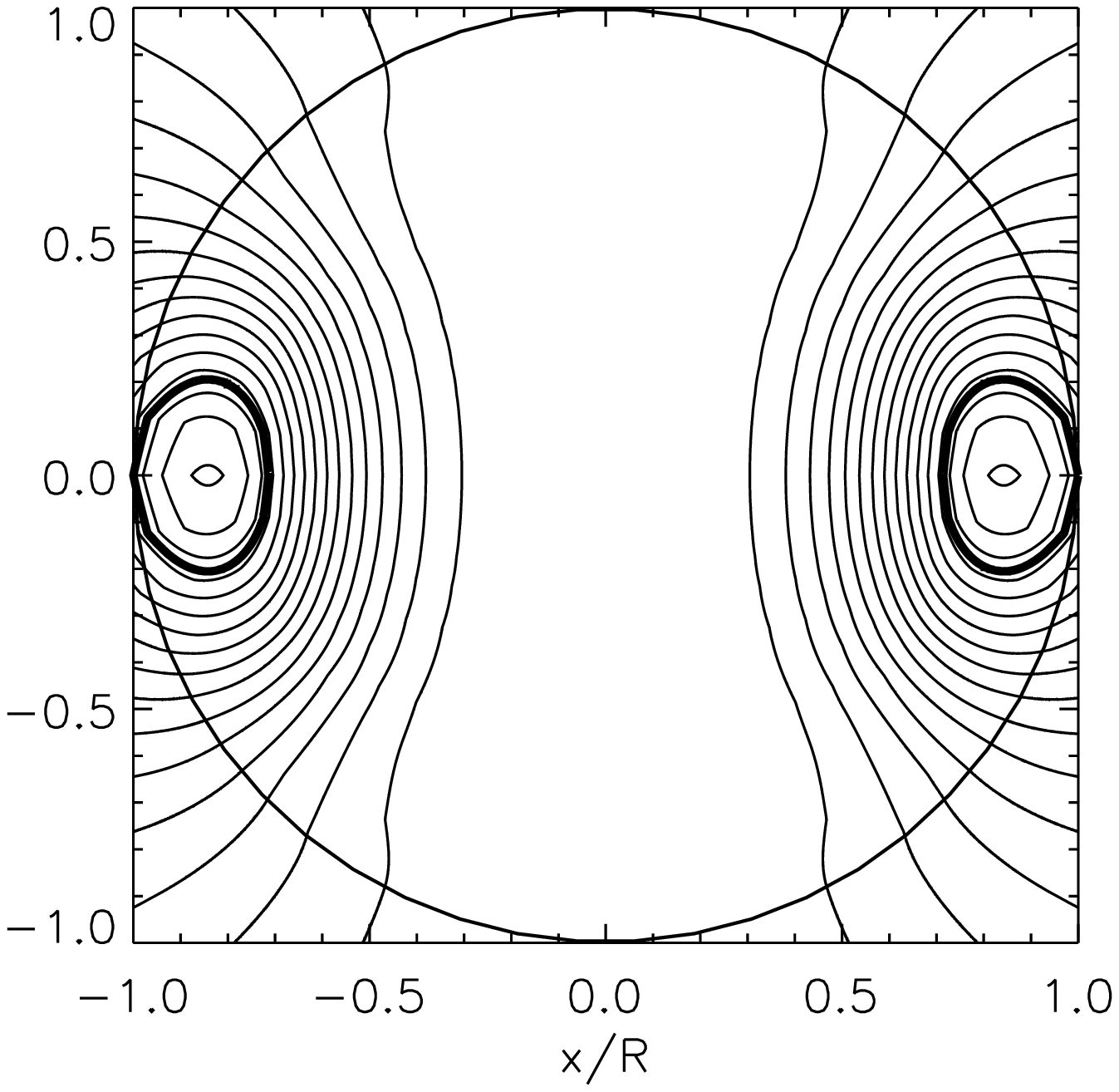}
\end{center} 
\caption{
The profiles of the tetrad components of the magnetic field
[$B_{(r)}(\theta=0)$, $B_{(\theta)}(\theta={\pi}/{2})$,
$B_{(\phi)}(\theta={\pi}/{2})$] are shown (upper panels).
In the lower panels we show the projection of
the field lines in the meridional plane.
Left and right panels refer, respectively, to the configuration ME1
and ME2. 
\label{Amin}}
\end{minipage} 
\end{figure*}
When we minimize the function $\delta M/H_m$ in the ME1 configuration, 
the $l>1$ multipoles, which were kept minimum in the MHM configuration, 
do not change $H_m$  significantly, but they change
$\delta M$, and since we require $\delta M$ to be minimum, they
combine as to reduce  the field in the inner region of the star.

\subsection{A more general choice of the function $\beta(\psi)$}\label{relation} 
In this section we construct  twisted torus configurations,
choosing two  different forms of
the function $\beta$, namely
\begin{equation}
\beta(\psi)=\psi \zeta_0 \left( |\psi/\bar{\psi}|-1 \right)^\sigma 
\Theta(|\psi/\bar{\psi}|-1)   \;\; ,
\label{zeta}
\end{equation}
(note that $\sigma=1$ corresponds to (\ref{zetaOLD})),
and 
\begin{equation}
\beta(\psi)= -\beta_0 \left( |\psi/\bar{\psi}|-1 \right)^\sigma 
\Theta(|\psi/\bar{\psi}|-1)   \;\; ,
\label{beta} 
\end{equation} 
where $\beta_0$ is a constant of order $O(B)$. A choice
similar to (\ref{zeta}) has been considered by Lander \& Jones
\shortcite{LJ}, who have studied the field configurations in a Newtonian framework.
Although Eqns. (\ref{zeta}), (\ref{beta}) do not exhaust 
all possible choices of the function $\beta(\psi)$,
 they are general enough to capture the main
features of the stationary twisted torus configurations.

For $\beta$ given by Eq. (\ref{zeta}) the magnetic field components and the GS
equation are 
\begin{eqnarray} B^\mu&=\Bigg(  &0 \;\; , \;\;
  \frac{e^{-\frac{\lambda}{2}}}{r^2\sin{\theta}} \psi_{,\theta} \;\; ,
  \;\; -\frac{e^{-\frac{\lambda}{2}}}{r^2\sin{\theta}}\psi_{,r} \;\; ,
  \nonumber\\
  &&-\frac{e^{-\frac{\nu}{2}} \zeta_0 \psi\left( |\psi/\bar{\psi}|-1
    \right)^\sigma}
  {r^2\sin^2{\theta}}~ \Theta(|\psi/\bar{\psi}|-1) \;\; \Bigg)
\label{Bfield1}
\end{eqnarray}
and 
\begin{eqnarray}
&&-\frac{e^{-\lambda}}{4\pi}\left[\psi''+\frac{\nu'-\lambda'}{2}\psi'
\right] -\frac{1}{4\pi r^2}\left[\psi_{,\theta\theta}-\cot{\theta}
\psi_{,\theta} \right] 
\nonumber\\
&&-\frac{e^{-\nu}\zeta_0^2}{4\pi}
\psi\left[\left(|\psi/\bar{\psi}|-1\right)^{2\sigma}+
\sigma |\psi/\bar{\psi}|\left(|\psi/\bar{\psi}|-1\right)^{2\sigma-1}
\right]\nonumber\\
&&\times\Theta(|\psi/\bar{\psi}|-1)
\nonumber\\ 
&&=(\rho+P)r^2\sin^2{\theta} [c_0+c_1\psi] \;\; .
\label{GS1}
\end{eqnarray}
For $\beta$ given by Eq. (\ref{beta}) they are:
\begin{eqnarray}
B^\mu&=\Bigg(
  &0 \;\; , \;\; \frac{e^{-\frac{\lambda}{2}}}{r^2\sin{\theta}}
  \psi_{,\theta} \;\; , \;\;
  -\frac{e^{-\frac{\lambda}{2}}}{r^2\sin{\theta}}\psi_{,r} \;\; ,
  \nonumber\\
  &&\frac{e^{-\frac{\nu}{2}} \beta_0 \left( |\psi/\bar{\psi}|-1
    \right)^\sigma}
  {r^2\sin^2{\theta}}~\Theta(|\psi/\bar{\psi}|-1) \;\; \Bigg) \;\;
  ,
\label{Bfield2}
\end{eqnarray}
and
\begin{eqnarray}
&&-\frac{e^{-\lambda}}{4\pi}\left[\psi''+\frac{\nu'-\lambda'}{2}\psi'
\right] -\frac{1}{4\pi r^2}\left[\psi_{,\theta\theta}-\cot{\theta}
\psi_{,\theta} \right] 
\nonumber\\
&&-\frac{e^{-\nu}\beta_0^2}{4\pi\psi}
\sigma |\psi/\bar{\psi}|\left(|\psi/\bar{\psi}|-1\right)^{2\sigma-1}
~\Theta(|\psi/\bar{\psi}|-1)
\nonumber\\ 
&&=(\rho+P)r^2\sin^2{\theta} [c_0+c_1\psi] \;\; .
\label{GS2}
\end{eqnarray}
\begin{figure}
\centering
\begin{center}
\includegraphics[width=4.7cm,angle=270]{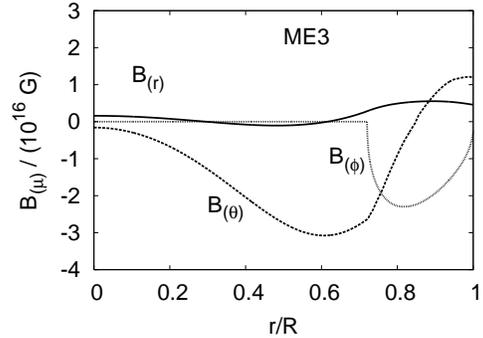}
\end{center}
\caption{The profiles of the tetrad components of the magnetic field
are shown for  the  configuration ME3, corresponding to
$\beta$ given by Eq.~(\ref{beta});
the values of the parameters are given in Eq. (\ref{conf2}).
\label{Bmin}}
\end{figure}
The field configurations 
are now identified by three parameters:
$(\sigma, c_1, \zeta_0)$ for the choice (\ref{zeta}), and $(\sigma, c_1,\beta_0)$ for the
choice (\ref{beta}). As in the previous section, we look for the minimal energy
configuration at fixed magnetic helicity; furthermore,  we compute the ratio of
the poloidal magnetic energy to the total magnetic energy.
We solve the system of GS equations for $l=1,3,5$ (they are given 
in appendix \ref{appGS} for both cases), with the boundary conditions
discussed in Section \ref{model}.
For each configuration we compute the magnetic helicity $H_m$, the correction
to the total energy $\delta M$, and the poloidal and toroidal
contributions to the magnetic energy $E_m$.  The equations to
determine $\delta M$ and $E_m$ are given in appendix \ref{HME}.  The
energetically favoured configurations are found by minimizing $\delta
M/H_m$ with respect to the three parameters.

Let us firstly consider the case in which the relation between toroidal
and poloidal fields is given by eq. (\ref{zeta}).
We find that the minimal energy configuration (for the APR2 EOS) corresponds to
\begin{eqnarray}
&&\sigma=0.18,\quad 
\zeta_0=0.20 ~\hbox{km}^{-1}, \nonumber\\
&&\sqrt{a_3^2(R)+a_5^2(R)}=3.4\times 10^{-3}~\hbox{km}\;.
\label{conf1}
\end{eqnarray}
We shall refer to this configuration as the ME2 configuration. In
Fig.~\ref{Amin} the configurations with $\sigma=1$ (ME1) and
$\sigma=0.18$ (ME2) are compared. In ME2 the magnetic field has a
slightly different shape: in particular, the toroidal component is
larger near the surface of the star, and the extension of the toroidal
field region along the $y$ axis is smaller. 
The ratio of the poloidal
magnetic energy to the total magnetic energy inside the star is
\begin{eqnarray}
&E_p/E_m=0.91 &\hbox{ for~ ME1} 
\nonumber\\
&E_p/E_m=0.87 &\hbox{ for ~ ME2}
\nonumber
\;\; .
\label{epem1}
\end{eqnarray}
If, as in Paper I, we include also the exterior field we find
$E_p/E_m=0.93$, $0.90$, respectively for the configurations ME1 and
ME2. Furthermore, we find that the minimal energy configuration is
nearly the configuration with smaller ratio $E_p/E_m$, i.e. with
larger toroidal component (confirming the results of Paper I): thus,
the $\sigma=0.18$ case also corresponds to the minimum value of
$E_p/E_m$ which can be obtained with the choice (\ref{zeta}).
We can conclude that if $\sigma$ is not assumed to be $1$, we can
obtain configurations with a larger toroidal contribution, but only by
a small amount.

We now consider the choice (\ref{beta}) for the function $\beta$.
In this case the  minimal energy configuration (for the APR2 EOS)
corresponds to
\begin{eqnarray}
&&\sigma=0.42,\quad 
\beta_0=9\times 10^{-4}, \nonumber\\
&&\sqrt{a_3^2(R)+a_5^2(R)}=3.7\times 10^{-3}~\hbox{km}\;.
\label{conf2}
\end{eqnarray}
This configuration, which is shown in Fig.~\ref{Bmin}, will be referred to
as the ME3 configuration. A comparison with the right panel of
Fig.~\ref{Amin} shows that the ME2 and ME3 configurations are very
similar. Inside the star $E_p/E_m=0.88$ and, as in the previous case,
it is the minimum value which can be obtained for this choice of the
function $\beta$.

For the GNH3 EOS, we obtain similar results. The minimal energy
configuration is obtained with the choice (\ref{zeta}), and
\begin{eqnarray}
&&\sigma=0.30,\quad 
\zeta_0=0.13 ~\hbox{km}^{-1},\nonumber\\
&&\sqrt{a_3^2(R)+a_5^2(R)}=5.1\times 10^{-3}~\hbox{km}\;.
\label{conf3}
\end{eqnarray}
The ratio of the poloidal energy to the total magnetic energy inside the star, 
for this configuration, is $E_p/E_m=0.93$.

We conclude that when we allow for a non-minimal contribution of 
the $l >1$ multipoles  and 
for a more general parametrization of the
function $\beta(\psi)$, the magnetic field changes
with respect to the MHM configuration found in Paper I as follows:
the poloidal field near the axis of the
star is smaller, and  the toroidal field near the stellar
surface is larger.  In all cases
the toroidal field never contributes to more than
$\sim13\%$ of the total magnetic energy stored inside the star.
\section{Structure deformations}\label{ell}

In this section we compute the quadrupole deformation induced by the
magnetic field for the twisted torus configurations previously
obtained. To this purpose we solve Einstein's
equations with a perturbative approach: the magnetic field, and the
deformations it induces, are considered as perturbations of a static,
spherically symmetric background. The relevant equations are described
in Appendix \ref{appell}.
The quantity which is relevant to estimate
the gravitational wave emission of the deformed star 
is the quadrupole ellipticity
\begin{equation}
\varepsilon_Q=\frac{Q}{I}
\;\; ,
\end{equation}
where $Q$ is the mass-energy quadrupole moment (see eq. (\ref{defQ})), 
and $I$ the mean value of the star's moment of inertia.  
Indeed, if the star rotates about an axis misaligned with the symmetry (or
magnetic) axis with a wobble angle $\alpha$,
it emits gravitational waves with amplitude
\cite{BG96}
\begin{equation}
  h_0\simeq \frac{4G}{rc^4}\Omega^2 I |\varepsilon_Q|\sin\alpha\label{h0eq}
  \;\; ,
\end{equation}
where $\Omega$ is the star angular velocity. 
We remind that we normalize the magnetic field
by fixing its value at the pole as $B_{pole}=10^{16}$ G (see Section
\ref{TTMFC}),  and that the quadrupole ellipticity  scales as 
$B_{pole}^2$.

Furthermore, it is interesting  to determine the sign of
the quadrupole ellipticity because, if $\varepsilon_Q<0$ (corresponding
to a prolate shape), a ``spin flip'' mechanism associated to viscous
forces may arise, as suggested in Jones \shortcite{PBJ}, Cutler
\shortcite{Cutler}. In this scenario, the angle between the magnetic
axis and the rotation axis would grow on a dissipation time-scale,
until they become orthogonal, and this process would be associated
to a large gravitational wave emission, potentially detectable by the
advanced generation of ground-based detectors Virgo and LIGO.

It is well known that while the poloidal field tends to make the star
oblate, the toroidal field tends to make it prolate.
Since in our configurations the
poloidal field dominates over the toroidal one, $\varepsilon_Q$ is always
positive. Therefore, twisted torus configurations are not compatible with the
spin-flip mechanism, and $|\varepsilon_Q|$ is larger for configurations in
which the toroidal contribution is smaller.
We find that, for the APR2 equation of state, $\varepsilon_Q=3.5\times
10^{-4}$ and $\varepsilon_Q=3.7\times 10^{-4}$ respectively for the ME2 and for the
ME3 configurations. 

It is also interesting to consider twisted torus configurations which do not
correspond to minimal energy. Indeed, it is not guaranteed that the
star sets in the minimal energy configuration before the crust forms
and the stellar matter become superfluid. Therefore, we have
determined the entire range of possible ellipticities for the twisted
torus configurations analyzed in this paper; we find $3.5\times
10^{-4}\lesssim\varepsilon_Q\lesssim4.8\times 10^{-4}$ for the APR2 EOS,
and $8.1\times 10^{-4}\lesssim\varepsilon_Q\lesssim9.6\times 10^{-4}$ for
the GNH3 EOS. As discussed above, the larger ellipticities are
obtained in the purely poloidal limit, whereas the smaller
refer to the minimal energy configurations. We
note that, as expected, given the mass
($1.4\,M_\odot$ in our case), less compact stars (GNH3) have larger
ellipticities. We also note that the values of $\varepsilon_Q$
we find  for the purely poloidal
case are similar to the maximal ellipticity found in Lander \& Jones
\shortcite{LJ}, where a polytropic EOS was employed.

Summarizing, the quadrupole ellipticity $\varepsilon_Q$ corresponding to 
$B_{pole}=10^{16}$ G would lie in the quite narrow ranges 
$(3.5,4.8)\times 10^{-4}$ for the APR2 EOS and $(8.1,9.6)\times 10^{-4}$ for the GNH3 EOS, 
i.e.
\begin{equation}
\varepsilon_Q\simeq k\left(\frac{B_{pole}[G]}{10^{16}}\right)^2\times 10^{-4}~,\label{finalformula}
\end{equation}
with $k\simeq4$ for the APR2 EOS and $k\simeq9$ for the GNH3 EOS.

\begin{figure}
\begin{center}
\centering
\includegraphics[width=4.7cm,angle=270]{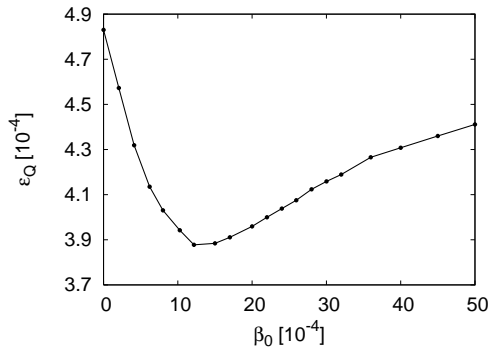}
\end{center} \caption{Ellipticities versus $\beta_0$ for $\sigma=1$,
 with $\beta(\psi)$ given by
 (\ref{beta}). }
\label{figell}
\end{figure}

As an example of the behaviour of $\varepsilon_Q$ when the toroidal field
contribution changes,
in Fig. \ref{figell} we plot $\varepsilon_Q$ versus the parameter $\beta_0$ 
for configurations obtained  choosing $\beta(\psi)$ as in Eq.
(\ref{beta}), assuming that  $\sigma=1$ and that the
contribution of the $l>1$ multipoles is fixed by energy minimization. 
In particular, we see that the 
maximal deformation is given by the purely poloidal configuration. 

\section{Concluding remarks}\label{conclusions}
In this paper we construct relativistic models of non-rotating,
stationary stars, with a twisted torus magnetic field configuration.
We extend the work done in Paper I, by removing the assumption of
minimal contribution from multipoles higher than $l=1$, by considering
more general forms of the function $\beta(\psi)$ which describes the
ratio between toroidal and  poloidal components, and by evaluating
the deformation that the magnetic field induces on the star.

We find that  the non-minimal contribution of the $l>1$
multipoles,   and the  more
general parametrization of the function
$\beta$, yield some interesting differences with
respect to the magnetic field configurations found in Paper I: the new
configurations have a much smaller poloidal field near the symmetry axis,
and a larger toroidal field near the stellar surface.
In any event,
the toroidal field never contributes to more
than $\sim13\%$ of the total magnetic energy stored inside the star.

Since the poloidal field always prevails, in the twisted torus configurations
the quadrupole ellipticity 
of the star $\varepsilon_Q$ is always positive, and its maximum value is
obtained in the  purely poloidal limit. 
As shown by eq. (\ref{finalformula}), 
which summarizes our results on the stellar deformation,
$\varepsilon_Q$ depends on the equation of
state of matter: less compact stars can have larger deformations.

We remark that the ellipticities given by eq. (\ref{finalformula}) are larger than
the bounds derived by evaluating the maximal strain that the crust can
sustain \cite{UCB00,Haskell06}. These bounds do not apply to the case
we study, because in our case the magnetic field is assumed to reach a stationary
configuration during the first few seconds of the neutron star life, when the star is still
fluid and no crust has formed yet; therefore, if the field is sufficiently
strong, the deformation it induces can be large,
and  may persist as the star cools down and the crust freezes in a
non-spherical shape
(see also the discussion in Haskell {\it et al.} \shortcite{Haskell},
Colaiuda {\it et al.} \shortcite{C08}).


Recent results from the  LIGO-Virgo collaboration \cite{LIGO08,LIGO09}
put an upper limit on the ellipticity of  the Crab pulsar, which
should be $\varepsilon_Q\lesssim 10^{-4}$. Our
study indicates that strongly magnetized neutron stars with ellipticities
of this order of magnitude  may exist, provided this strong deformation
was built up before the crust was formed, and the magnetic 
field was sufficiently strong.

In order to further substantiate this scenario, an 
important issue which remains  to be clarified 
is whether the twisted torus configurations
we find are stable.  This issue will be the subject of a future
investigation. 

\section*{Acknowledgments}\label{acknow} 
We thank Jos\'e Pons, Luigi Stella, Cristiano Palomba and Ulrich Geppert for useful
suggestions and discussions.

This work was partially supported by CompStar, a Research Networking
Program of the European Science Foundation. L.G. has been partially
supported by the grant PTDC/FIS/098025/2008.

\appendix 
\section{GS equations}\label{appGS} 
The harmonic expansion of the GS equations (\ref{GS1}), corresponding
to the choice (\ref{zeta}) of the function $\beta(\psi)$, gives (if we
include the $l=1,3,5$ components in the expansion)
\begin{eqnarray}
&&\frac{1}{4\pi}\left(e^{-\lambda}a_1''
+e^{-\lambda}\frac{\nu'-\lambda'}{2}a_1'-\frac{2}{r^2}a_1 \right)
\nonumber\\
&&-\frac{e^{-\nu}}{4\pi} \int_0^\pi (3/4) 
\; \zeta_0^2 \psi\left[
\left(|\psi/\bar{\psi}|-1\right)^{2\sigma}\right.\nonumber\\
&&\left.+
\sigma |\psi/\bar{\psi}|\left(|\psi/\bar{\psi}|-1\right)^{2\sigma-1}
\right]~\Theta(|\psi/\bar{\psi}|-1)\sin{\theta}\;d\theta\nonumber\\
&&=\left[c_0-\frac{4}{5}c_1\left(a_1-\frac{3}{7}a_3\right)\right] 
(\rho+P)r^2\,,  \label{firstL1}
\end{eqnarray}
\begin{eqnarray}
&&\frac{1}{4\pi}\left(e^{-\lambda}a_3''
+e^{-\lambda}\frac{\nu'-\lambda'}{2}a_3'-\frac{12}{r^2}a_3 \right)
\nonumber\\
&&+\frac{e^{-\nu}}{4\pi} \int_0^\pi (7/48) 
\; \zeta_0^2 \psi\left[
\left(|\psi/\bar{\psi}|-1\right)^{2\sigma}\right. \nonumber\\
&&\left.+
\sigma |\psi/\bar{\psi}|\left(|\psi/\bar{\psi}|-1\right)^{2\sigma-1}
\right]\nonumber\\
&&\times \Theta(|\psi/\bar{\psi}|-1)(3-15\cos^2{\theta})
\sin{\theta}\;d\theta\nonumber\\
&&=c_1 (\rho+P)r^2 \left(\frac{2}{15}a_1-\frac{8}{15}a_3+
\frac{10}{33}a_5\right)  \;\; ,
\label{firstL3}
\end{eqnarray}
\begin{eqnarray}
&&\frac{1}{4\pi}\left(e^{-\lambda}a_5''
+e^{-\lambda}\frac{\nu'-\lambda'}{2}a_5'-\frac{30}{r^2}a_5 \right)
\nonumber\\
&&+\frac{e^{-\nu}}{4\pi} \int_0^\pi (11/60) 
\; \zeta_0^2 \psi\left[
\left(|\psi/\bar{\psi}|-1\right)^{2\sigma}\right.\nonumber\\
&&\left.+
\sigma |\psi/\bar{\psi}|\left(|\psi/\bar{\psi}|-1\right)^{2\sigma-1}
\right]\nonumber\\
&&\times\Theta(|\psi/\bar{\psi}|-1)
\frac{(-315\cos^4{\theta}+210\cos^2{\theta}-15)}{8}
\sin{\theta}\;d\theta\nonumber\\
&&
=c_1 (\rho+P)r^2 \left(\frac{4}{21}a_3-\frac{20}{39}a_5\right)  \;\; ,
\label{firstL5}
\end{eqnarray}
where
\begin{eqnarray}
\psi&=&\left[-a_1+\frac{a_3 (3-15\cos^2{\theta})}{2}\right.
\nonumber\\
&&\left.+\frac{a_5 (-315\cos^4{\theta}+210\cos^2{\theta}-15)}{8}\right] 
\sin^2{\theta}  \;\; .  
\label{psi135}
\end{eqnarray}
The harmonic expansion of the GS equations (\ref{GS2}), corresponding
to the choice (\ref{beta}) of the function $\beta(\psi)$, gives 
\begin{eqnarray}
&&\frac{1}{4\pi}\left(e^{-\lambda}a_1''
+e^{-\lambda}\frac{\nu'-\lambda'}{2}a_1'-\frac{2}{r^2}a_1 \right)
\nonumber\\
&&-\frac{e^{-\nu}}{4\pi} \int_0^\pi (3/4) 
\; \frac{\beta_0^2}{\psi}\sigma |\psi/\bar{\psi}|\left(|\psi/\bar{\psi}|-1\right)^{2\sigma-1}
\nonumber\\
&&\times\Theta(|\psi/\bar{\psi}|-1)\sin{\theta}\;d\theta\nonumber\\
&&=\left[c_0-\frac{4}{5}c_1\left(a_1-\frac{3}{7}a_3\right)\right] 
(\rho+P)r^2\,,  \label{secondL1}
\end{eqnarray}
\begin{eqnarray}
&&\frac{1}{4\pi}\left(e^{-\lambda}a_3''
+e^{-\lambda}\frac{\nu'-\lambda'}{2}a_3'-\frac{12}{r^2}a_3 \right)
\nonumber\\
&&+\frac{e^{-\nu}}{4\pi} \int_0^\pi (7/48) 
\; \frac{\beta_0^2}{\psi}\sigma |\psi/\bar{\psi}|\left(|\psi/\bar{\psi}|-1\right)^{2\sigma-1}
\nonumber\\
&&\times \Theta(|\psi/\bar{\psi}|-1)(3-15\cos^2{\theta})
\sin{\theta}\;d\theta\nonumber\\
&&=c_1 (\rho+P)r^2 \left(\frac{2}{15}a_1-\frac{8}{15}a_3+
\frac{10}{33}a_5\right)  \;\; ,
\label{secondL3}
\end{eqnarray}
\begin{eqnarray}
&&\frac{1}{4\pi}\left(e^{-\lambda}a_5''
+e^{-\lambda}\frac{\nu'-\lambda'}{2}a_5'-\frac{30}{r^2}a_5 \right)
\nonumber\\
&&+\frac{e^{-\nu}}{4\pi} \int_0^\pi (11/60) 
\; \frac{\beta_0^2}{\psi}\sigma |\psi/\bar{\psi}|\left(|\psi/\bar{\psi}|-1\right)^{2\sigma-1}
\nonumber\\
&&\times\Theta(|\psi/\bar{\psi}|-1)
\frac{(-315\cos^4{\theta}+210\cos^2{\theta}-15)}{8}
\sin{\theta}\;d\theta\nonumber\\
&&
=c_1 (\rho+P)r^2 \left(\frac{4}{21}a_3-\frac{20}{39}a_5\right)  \;\; ,
\label{secondL5}
\end{eqnarray}
where $\psi$ is the same as in (\ref{psi135}).

\section{Energy and magnetic helicity}\label{HME} 

The magnetic helicity is given by eq. (\ref{firstHm})
where, if $\beta(\psi)$ is given by eq. (\ref{zeta}), $A_{r,\theta}$ and $A_r$ are
\begin{eqnarray} 
A_{r,\theta}&=& \frac{e^{\frac{\lambda-\nu}{2}}}{\sin{\theta}} 
\psi \zeta_0 \left( |\psi/\bar{\psi}|-1 \right)^{\sigma}
\Theta(|\psi/\bar{\psi}|-1)
 \;\; ,\nonumber\\ 
A_r&=& e^{\frac{\lambda-\nu}{2}}\zeta_0 
\int_0^\theta \frac{\psi}{\sin{\theta'}}
\left( |\psi/\bar{\psi}|-1 \right)^{\sigma}
\nonumber\\
&&\qquad\qquad\qquad\qquad\times\Theta(|\psi/\bar{\psi}|-1) d\theta' 
  \;.
\label{firstHmbis}
\end{eqnarray}  
The total energy of the system is  $E=M+\delta M$, where $M$
is the mass of the (spherically symmetric) star without magnetic
field and   $\delta M$ is the contribution induced by the magnetic field.
It can be determined by considering the far field limit of the spacetime
metric \cite{MTW,Thorne}, in terms of the
function $m_0(r)$ defined in equation (\ref{defmetric}):
\begin{equation}
\delta M=\lim_{r\rightarrow \infty}m_0(r)
  \;\; . 
\label{limHm}
\end{equation}
The components of the perturbed Einstein's equations relevant for the
determination of $m_0(r)$ give, for the choice (\ref{zeta}) of the
function $\beta(\psi)$,
\begin{eqnarray}
&&m'_0-4\pi r^2\frac{\rho'}{P'}\delta
p_0=\frac{1}{3}(a'_1)^2 e^{-\lambda} +\frac{6}{7}(a'_3)^2
e^{-\lambda}\nonumber\\
&&+\frac{15}{11}(a'_5)^2 e^{-\lambda}
+\frac{2}{3r^2}a_1^2+\frac{72}{7r^2}a_3^2
+\frac{450}{11r^2}a_5^2\nonumber\\
&&+\frac{e^{-\nu}}{4}\left[
\int_0^\pi \zeta_0^2 \left( |\psi/\bar{\psi}|-1 \right)^{2\sigma}
\Theta(|\psi/\bar{\psi}|-1) \frac{\psi^2}{\sin{\theta}} d\theta\right]
\;\; , \nonumber\\ 
&&\delta p'_0+\left[\frac{\nu'}{2}\left( \frac{\rho'}{P'}+1 \right)+4\pi r
e^\lambda (\rho+P)\right]\delta p_0 \nonumber\\
&&+e^{2\lambda}m_0(\rho+P)\left(
\frac{1}{r^2}+8\pi P\right)\nonumber\\ 
&&=(\rho+P)\Bigg\{-\frac{2}{3}a_1'\left[c_0-\frac{4}{5}c_1 \left( a_1-
\frac{3}{7}a_3\right) \right] \nonumber\\
&&-\frac{12}{7}a_3'c_1 \left(
\frac{2}{15}a_1-\frac{8}{15}a_3+\frac{10}{33}a_5 \right) \nonumber\\
&&-\frac{10}{11}a_5'c_1 \left(
\frac{4}{21}a_3-\frac{20}{39}a_5 \right)
-\frac{1}{3r}(a'_1)^2-\frac{6}{7r}(a'_3)^2 \nonumber\\
&&-\frac{15}{11r}(a'_5)^2+\frac{2e^{\lambda}}{3r^3}a_1^2 
+\frac{72e^{\lambda}}{7r^3}a_3^2+\frac{450e^{\lambda}}{11r^3}a_5^2
\nonumber\\
&&-\frac{e^{\lambda-\nu}}{4r}\left[ \int_0^\pi \zeta_0^2 \left(
|\psi/\bar{\psi}|-1 \right)^{2\sigma} \Theta(|\psi/\bar{\psi}|-1)
\frac{\psi^2}{\sin{\theta}} d\theta \right] \Bigg\} \nonumber\\
\label{firsteqmp}
\end{eqnarray}
(where $\delta p_0$ is the $l=0$ component of the pressure perturbation).

Finally, the magnetic energy is \cite{Straumann,C09}
\begin{equation}
E_m=\frac{1}{2}\int_0^\infty r^2 e^{\frac{\lambda+\nu}{2}}dr\int_0^\pi 
\sin{\theta}B^2 d\theta \;\; . 
\label{Em}
\end{equation}
The above formula can be used to compute the relative amount of magnetic 
energy associated with toroidal and poloidal fields. 

The equations for the choice (\ref{beta}) of the function $\beta$
can be obtained  by the following 
substitutions: \\$\psi\zeta_0\rightarrow -\beta_0$ in (\ref{firstHmbis}); 
$\psi^2\zeta_0^2\rightarrow \beta_0^2$ in (\ref{firsteqmp}). 

\section{Quadrupole deformations}\label{appell}

The perturbed metric can be written as \cite{IS,C08}
\begin{eqnarray}
&&ds^2=-e^{\nu}\Big( 1+2[h_0(r)+h_2(r) P_2(\cos{\theta})]\Big) dt^2
\nonumber\\
&&+2\Big[ i_1(r) P_1(\cos{\theta})
+i_2(r) P_2(\cos{\theta})+i_3(r) P_3(\cos{\theta}) \Big] dtdr
\nonumber\\
&&+2\sin{\theta}\left(
v_1\frac{\partial}{\partial \theta}P_1(\cos{\theta})+
v_2\frac{\partial}{\partial \theta}P_2(\cos{\theta})\right. \nonumber\\
&&\left.+
v_3\frac{\partial}{\partial \theta}P_3(\cos{\theta})
\right)dtd\phi
\nonumber\\
&&+2\sin{\theta}\left(
w_2\frac{\partial}{\partial \theta}P_2(\cos{\theta})+
w_3\frac{\partial}{\partial \theta}P_3(\cos{\theta})
\right)drd\phi
\nonumber\\
&&+e^{\lambda}\biggl[1+\frac{2e^{\lambda}}{r}
\Big( m_0(r)+m_2(r) P_2(\cos{\theta}) \Big)
\biggr]dr^2 
\nonumber\\
&&+r^2\Big[1+2 k_2(r)P_2(\cos{\theta})\Big]\biggl(d\theta^2+sin^2 \theta d\phi^2\biggr)
  \;\; .
\label{defmetric}
\end{eqnarray}

The quadrupole ellipticity of the star is defined as $\varepsilon_Q=Q/I$
\footnote{Note that in Colaiuda {\it et al.} \shortcite{C08} eq. (81) 
has a wrong minus sign.},
where $Q$ is the mass-energy quadrupole moment, given by the far field
limit of the metric (\ref{defmetric}) 
\begin{equation}
h_2(r\rightarrow \infty)\sim
Q/r^3 \;\;, 
\label{defQ}
\end{equation}
and $I$ is the mean value of the moment of inertia of the
star. The value of $I$ can be estimated from the limit
$\Omega\rightarrow 0$ of the ratio $J/\Omega$ in a slowly rotating
star model ($\Omega$ is the angular velocity, $J$ the angular
momentum).  For $M=1.4$ $M_\odot$ we have $I=98.39$ km$^3$ (APR2 EOS)
and $I=134.6$ km$^3$ (GNH3 EOS).

In order to compute $\varepsilon_Q$ we need to solve the following system 
of linearized Einstein equations (here we consider the choice (\ref{zeta})):
\begin{eqnarray}
&&k'_2+h'_2-h_2\biggl(\frac{1}{r}-\frac{\nu'}{2}\biggr)
-m_2\biggl(\frac{\nu'}{2}+\frac{1}{r}\biggr)\frac{e^{\lambda}}{r}
\nonumber\\
&&=\frac{5}{4r^2}\int_0^\pi \psi_{,\theta}\psi_{,r}
\frac{(3\cos^2{\theta}-1)\cot{\theta}}{\sin{\theta}}d\theta
  \;\; ,
\label{eqrth}\\
&& h_2+\frac{e^{\lambda}}{r}m_2 \nonumber\\
&&=\frac{5}{4r^2}\int_0^\pi \bigg(
-(\psi_{,r})^2 r^2 e^{-\lambda}
+e^{-\nu}\zeta_0^2 \psi^2 r^2 (|\psi/\bar{\psi}|-1)^{2\sigma}\nonumber\\
&&\times \Theta(|\psi/\bar{\psi}|-1) 
\bigg)~\frac{(3\cos^2{\theta}-1)}{\sin{\theta}}d\theta 
  \;\; , 
\label{eqphph}\\
&&\left( \nu'+\frac{2}{r}\right)k'_2+\frac{2}{r}h'_2
-\frac{4}{r^2}e^{\lambda}k_2-\frac{6}{r^2}e^{\lambda}h_2\nonumber\\
&&-\left( \frac{1}{r^2}+8\pi P\right)\frac{2e^{2\lambda}}{r} m_2
-8\pi e^{\lambda}\delta p_2
\nonumber\\
&&=\frac{5}{4r^4}e^{\lambda}\int_0^\pi \bigg(
-(\psi_{,\theta})^2+(\psi_{,r})^2 r^2 e^{-\lambda}\nonumber\\
&&+e^{-\nu}\zeta_0^2 \psi^2 r^2 (|\psi/\bar{\psi}|-1)^{2\sigma} \Theta(|\psi/\bar{\psi}|-1) 
\bigg) \nonumber\\
&&\times\frac{(3\cos^2{\theta}-1)}{\sin{\theta}}d\theta   \;\; ,\label{eqrr}
\end{eqnarray}
where
\begin{eqnarray}
\psi&=&\left[-a_1+\frac{a_3 (3-15\cos^2{\theta})}{2}\right.\nonumber\\
&&\left.+\frac{a_5 (-315\cos^4{\theta}+210\cos^2{\theta}-15)}{8}\right] 
\sin^2{\theta}  \;\; .  
\nonumber
\end{eqnarray}
The integration can be simplified by introducing the auxiliary function
\begin{equation}
y_2=k_2+h_2+W(r,\theta)  \;\; ,
\end{equation} 
where
\begin{eqnarray}
W(r,\theta)&=&\frac{5e^{-\lambda}}{16r^2}\int_0^\pi \bigg(
-e^{\lambda}(\psi_{,\theta})^2+r^2 (\psi_{,r})^2  \nonumber\\
&&-2r\psi_{,\theta}\psi_{,r}\cot{\theta}\bigg)
~\frac{(3\cos^2{\theta}-1)}{\sin{\theta}}d\theta
  \;\; . 
\end{eqnarray} 
This generalizes the variable change adopted in Ioka \& Sasaki \shortcite{IS},
Colaiuda {\it et al.} \shortcite{C08}.
With the above substitution we are left with two coupled equations
\begin{eqnarray} 
&&\quad y'_2+\nu'h_2=W'+\frac{5}{4 r^2}  
\int_0^\pi \bigg[ \psi_{,\theta}\psi_{,r}\cot{\theta}+\left( \frac{\nu'}{2}+\frac{1}{r}\right)\nonumber\\
&& \times \bigg(-(\psi_{,r})^2 r^2 e^{-\lambda}
+e^{-\nu}\zeta_0^2 \psi^2 r^2 (|\psi/\bar{\psi}|-1)^{2\sigma}\nonumber\\
&& \times\Theta(|\psi/\bar{\psi}|-1) 
\bigg)\bigg]\frac{(3\cos^2{\theta}-1)}{\sin{\theta}}d\theta 
  \;\; , \label{eqAA}\\ 
&&\quad h'_2\!+\!\frac{4}{\nu' r^2}e^{\lambda}y_2 
\!+\!\left[\nu'\!-\!\frac{8\pi e^{\lambda}}{\nu'}(\rho\!+\!P)\!+\!
\frac{2}{\nu' r^2}(e^{\lambda}\!-\!1) 
\right]h_2 \nonumber\\ 
&&\quad \quad=\frac{5}{8r^2}\int_0^\pi \bigg[ 
-\nu' e^{-\lambda}r^2(\psi_{,r})^2+2 \psi_{,r}\psi_{,\theta}\cot{\theta}
\nonumber\\ 
&&\quad \quad+e^{-\nu}\zeta_0^2\psi^2 \left( \nu'r^2-\frac{2}{\nu'}e^\lambda\right)
(|\psi/\bar{\psi}|-1)^{2\sigma}\nonumber\\
&&\times \Theta(|\psi/\bar{\psi}|-1)\bigg] \frac{(3\cos^2{\theta}-1)}{\sin{\theta}}d\theta 
\nonumber\\
&&+\frac{10\pi}{\nu'} e^{\lambda}(\rho+P)\int_0^\pi 
[c_0+c_1\psi]\psi_{,\theta} \sin^2{\theta}\cos{\theta} d\theta
  \;\; ,\label{eqBB}
\end{eqnarray}
where we have used the following relation (arising from $T^{\theta\nu}_{\;\; ;\nu}=0$):
\begin{eqnarray}
\delta p_2&=&-(\rho+P)\left(h_2+\frac{5}{4}\int_0^\pi [c_0+c_1\psi]\psi_{,\theta} 
\sin^2{\theta}\cos{\theta} d\theta \right)
  \;\; .\nonumber\\
\end{eqnarray}
Eqns. (\ref{eqAA}), (\ref{eqBB}) can be solved using the same 
procedure described in Colaiuda {\it et al.} \shortcite{C08}.
If we adopt the choice (\ref{beta}) for the
relation between toroidal and poloidal fields, we proceed in the same way. In this case the final system of equations
writes
\begin{eqnarray}
&&\quad y'_2+\nu'h_2=W'+\frac{5}{4 r^2}  
\int_0^\pi \bigg[ \psi_{,\theta}\psi_{,r}\cot{\theta}+\left( \frac{\nu'}{2}+\frac{1}{r}\right)\nonumber\\
&& \times \bigg(-(\psi_{,r})^2 r^2 e^{-\lambda}
+e^{-\nu}\beta_0^2 r^2 (|\psi/\bar{\psi}|-1)^{2\sigma}\nonumber\\
&&\times \Theta(|\psi/\bar{\psi}|-1) 
\bigg)\bigg]\frac{(3\cos^2{\theta}-1)}{\sin{\theta}}d\theta 
  \;\; , \\ 
&&\quad h'_2\!+\!\frac{4}{\nu' r^2}e^{\lambda}y_2 
\!+\!\left[\nu'\!-\!\frac{8\pi e^{\lambda}}{\nu'}(\rho\!+\!P)\!+\!\frac{2}{\nu' r^2}(e^{\lambda}\!-\!1) 
\right]h_2 \nonumber\\ 
&&\quad=\frac{5}{8r^2}\int_0^\pi \bigg[ 
-\nu' e^{-\lambda}r^2(\psi_{,r})^2+2 \psi_{,r}\psi_{,\theta}\cot{\theta}
\nonumber\\ 
&&\quad+e^{-\nu}\beta_0^2 \left( \nu'r^2-\frac{2}{\nu'}e^\lambda\right)
(|\psi/\bar{\psi}|-1)^{2\sigma} \nonumber\\
&&\times \Theta(|\psi/\bar{\psi}|-1)\bigg]
\frac{(3\cos^2{\theta}-1)}{\sin{\theta}}d\theta 
\nonumber\\ 
&&+\frac{10\pi}{\nu'} e^{\lambda}(\rho+P)\int_0^\pi 
[c_0+c_1\psi]\psi_{,\theta} \sin^2{\theta}\cos{\theta} d\theta
  \;\; .
\end{eqnarray}



\end{document}